\documentclass[authoryear,11pt]{article}
\usepackage{url}
\usepackage{natbib}
\usepackage[english]{babel}
\usepackage{amsmath}
\usepackage{latexsym}
\usepackage{amssymb}
\usepackage{graphicx}
\usepackage{subfigure}
\usepackage{color}

\usepackage{comment}
\includecomment{figurepdf}

\newcommand{\G}{\mathcal{G}}

\newcommand{\E}{\mathcal{E}}

\newcommand{\Pt}{\pi}

\newcommand{\Ptxy}{\Pi}

\newcommand{\lp}{\langle}
\newcommand{\rp}{\rangle}

\newcommand{\edge}{-\!\!\!-{}}

\newcommand{\given}{\cdot}

\newcommand{\vc}[2]{(#1)(#2)}

\newcommand{\cvc}[3]{(#1)(#2)\given#3}

\newcommand{\w}{\omega}

\newcommand{\ew}{\omega}

\newcommand{\ewnorm}{\psi}

\newcommand{\ewsub}[2]{#1\given(#2)}

\let\Sigma\varSigma
\let\Gamma\varGamma
\let\Theta\varTheta

\newcommand{\Kita}{\mathit{K}}

\newcommand{\Xbf}{\mathbf{X}}

\usepackage{amsthm}
\theoremstyle{plain}
\newtheorem{thm}{Theorem}[section]

\theoremstyle{definition}

\newtheorem{exmp}{Example}[section]

\theoremstyle{remark}

\begin{document}
\title{The Networked Partial Correlation and its Application to the Analysis of Genetic Interactions}

\author{
Alberto Roverato\\ Universit\`a di Bologna, Italy\\
\texttt{alberto.roverato@unibo.it}
\and
Robert Castelo\\ Universitat Pompeu Fabra, Barcelona, Spain\\ \texttt{robert.castelo@upf.edu}
}
\date{\today}
\maketitle
\begin{abstract}\small
Genetic interactions confer robustness on cells in response to genetic perturbations. This often occurs through molecular buffering mechanisms that can be predicted using, among other features, the degree of coexpression between genes, commonly estimated through marginal measures of association such as Pearson or Spearman correlation coefficients. However, marginal correlations are sensitive to indirect effects and often partial correlations are used instead. Yet, partial correlations convey no information about the (linear) influence of the coexpressed genes on the entire multivariate system, which may be crucial to discriminate functional associations from genetic interactions. To address these two shortcomings, here we propose to use the edge weight derived from the covariance decomposition over the paths of the associated gene network. We call this new quantity the networked partial correlation and use it to analyze genetic interactions in yeast.
\end{abstract}


\section{Introduction}\label{SEC:introduction}
The deletion of individual genes in model organisms, such as the budding yeast,
\textit{Saccharomyces cerevisiae}, produces mutants that cannot express
the knocked-out gene, constituting one of the primary tools in experimental
genetics to elucidate gene function. However, the systematic culture of yeast
single-gene mutants has revealed that the majority of its about 6000 genes are
dispensable because no sizable effect in fitness can be observed among the
corresponding mutants \citep{winzeler1999functional}. An explanation to this
observation is the presence of buffering relationships between pairs of genes,
by which the absence of one gene is counterbalanced by the expression of its
partner. The simultaneous deletion of two genes produces a so-called
\textit{double mutant} organism. When the change in fitness of a double-mutant
significantly deviates from the expected change resulting from the combination
of the two single mutant fitness effects, then one concludes that there is a
so-called \textit{genetic interaction} between these two genes. A reduction in
fitness by a double-mutant is known as \textit{synthetic sickness} and the
extreme case of this phenomenon, which is known as \textit{synthetic lethality},
occurs when two single mutants are still viable but the genetic interaction of
the two knocked-out genes leads to cell death \citep{tucker2003lethal}. This
concept has been exploited in the field of cancer research to tackle the
resistance to chemotherapeutics by trying to target multiple oncogenes
simultaneously \citep{luo2009genome,jerby2014predicting}.

Genetic interactions can be experimentally identified in a number of ways. One
of them consists of measuring the deviation in fitness between the expected
effect of combining two single mutants and the observed effect of the
corresponding double mutant \citep{baryshnikova2010quantitative}.
Yet, producing an exhaustive catalogue of single and double mutants to enable
the exploration of all possible genetic interactions is only feasible in model
organisms with a moderate number of genes, such as yeast. For this reason,
it is important to have computational tools that enable predicting genetic
interactions in larger model organisms \citep{eddy2006total} and, ideally, in
humans \citep{deshpande2013comparative} where, in addition to the reduced
possibilities for genetic manipulation, the number of possible gene pairs can
be tenfold larger.

The simultaneous expression of two genes, known as \textit{gene coexpression},
is a proxy for the presence of a functional association between them. The
high-throughput profiling of expression for thousands of genes in parallel
provides multivariate data whose analysis with clustering techniques and
graphical models has proven to be useful for exploring gene coexpression in
terms of gene network representations of the data. This has been exploited in
a number of applications ranging from inferring function in poorly characterized
genes to predicting buffering relationships behind genetic interactions
\citep{eisen1998cluster,friedman2004inferring,wong2004combining,jerby2014predicting}.
Existing approaches that attempt to predict genetic interactions not only use
gene coexpression but also many other biological features such as protein
function and localization, homology relationships and protein-protein
interactions \citep{wong2004combining,zhong2006genome,conde2009human,deshpande2013comparative,jerby2014predicting},
which we will not consider in this paper.

Gene coexpression is commonly identified using Pearson or Spearman correlation
coefficients. However, the marginal nature of these quantities often leads to
spurious associations resulting from indirect effects and nonbiological sources
of variation. To address this problem, we can use graphical Gaussian models
\citep{dempster1972covariance,whittaker1990graphical} in which a key role is
played by the partial covariance because if a pair of variables is not joined
by an edge in the network, then the corresponding partial covariance is equal to zero. The partial covariance can be normalized
to obtain a partial correlation that, in the molecular context, can be regarded
as the natural measure of the strength of the direct association between the two
genes forming an edge in the network \citep{delafuente2004discovery,castelo2006robust,zuo2014biological}.
However, although partial correlation is a measure of direct coexpression
between genes, the buffering mechanism behind a genetic interaction not only
leads to gene coexpression but also confers robustness on the whole system in
response to genetic perturbations \citep{nijman2011synthetic}. From this
perspective, the information that is provided by the value of a partial
correlation is not sufficient to capture such a robustness, reflected in the
functional relationship between the two intervening genes and the remaining
genes in the system.

One of the first attempts to describe the influence of a direct association
within an entire multivariate system was provided by \citet{wright1921correlation},
who described the covariance decomposition between two variables along their
connecting paths in a directed graph \citep[see also][for a recent review]{chen2015graphical}.
More recently, within the analysis of undirected graphical Gaussian models,
\citet{jones2005covariance} showed how the covariance between two variables
can be computed as the sum of weights associated with the undirected paths
joining the variables, providing the undirected counterpart to the results
of \citet{wright1921correlation}. In this paper, we make the observation that
every single edge in a network can be regarded as a path, and therefore, can
also have such a weight associated with it. We investigate how that weight
captures both the strength of the direct association between the two variables
and their relationship with the remaining variables in the system. We provide an
interpretation of these edge weights that suggests us to name them
\emph{networked partial covariances} and then we normalize them to obtain
\emph{networked partial correlations}.
We demonstrate how the covariance turns out to be a special case of the networked
partial covariance, and how this result generalizes the covariance decomposition
of \citet{jones2005covariance}. Finally, we demonstrate how networked partial
correlations improve marginal and partial correlations as proxies for the
presence of buffering relationships behind genetic interactions in yeast.

This paper is organized as follows. Section~\ref{SEC:notation} provides the
required background on undirected graphical models, path weights and the partial
vector correlation coefficient. In Section~\ref{SEC:NPC} the definitions of the
networked partial covariance and correlation are given and their interpretation
discussed. The limited-order networked partial covariance and its decomposition
are given in Section~\ref{SEC:limited.order.NPC.dec}.
Section~\ref{SEC:genetic.interactions} presents the results on the
analysis of genetic interactions in yeast whereas in the Supplementary Material
we provide full details on how the data were analysed. Finally,
Section~\ref{SEC:discussion} contains a discussion.

\section{Notation and background}\label{SEC:notation}
\subsection{Undirected graphical models}
Let $\Xbf\equiv \Xbf_{V}$ be a random vector indexed by a finite set $V=\{1,\ldots,p\}$ so that for $A\subseteq V$, $\Xbf_{A}$ is the subvector of $\Xbf$ indexed by $A$. The random vector $\Xbf_{V}$ has probability distribution $\mathbb{P}_{V}$ and we denote the covariance matrix of $\Xbf_{V}$ by $\Sigma=\Sigma_{VV}=\{\sigma_{uv}\}_{u,v\in V}$ and the \emph{concentration} (or precision) matrix by $\Sigma^{-1}=\Kita=\{\kappa_{uv}\}_{u,v\in V}$. For $B\subseteq V$ with $A\cap B=\emptyset$ the \emph{partial covariance} matrix $\Sigma_{AA\given B}=\Sigma_{AA}-\Sigma_{AB}\Sigma_{BB}^{-1}\Sigma_{BA}$ is the covariance matrix of $\Xbf_{A}|\Xbf_{B}$, that is the residual vector deriving from the linear least square predictor of $\Xbf_{A}$ from $\Xbf_{B}$ \citep[see][p.~134]{whittaker1990graphical}. Recall that, in the Gaussian case,  $\Sigma_{AA\given B}$ coincides with the covariance matrix of the conditional distribution of $\Xbf_{A}$ given $\Xbf_{B}$. We use the convention that we write $\Sigma_{AA}^{-1}$ when the submatrix extraction is performed before the inversion, that is $\Sigma_{AA}^{-1}=(\Sigma_{AA})^{-1}$ and, similarly, $\Sigma_{AA\given B}^{-1}=(\Sigma_{AA\given B})^{-1}$. We write $\bar{A}=V\backslash A$ to denote the complement of a subset $A$ with respect to $V$ and recall that, from the rule for the inversion of a partitioned matrix, $\Sigma_{AA\given \bar{A}}^{-1}=\Kita_{AA}$ and, accordingly, $\Sigma_{AA}^{-1}=\Kita_{AA\given \bar{A}}$.

An \emph{undirected graph} with vertex set $V$ is a pair $\G=(V, \E)$ where $\E$ is a set of \emph{edges}, which are unordered pairs of vertices; formally $\E\subseteq V\times V$. The graphs we consider have no self-loops, that is $\{v, v\}\not\in \E$ for any $v\in V$. The subgraph of $\G$ \emph{induced} by $A\subseteq V$ is the undirected graph $\G_{A}$ with vertex set $A$ and edges $\E_{A}=\{\{u, v\}\in \E: u,v\in A\}$. A \emph{path} between $x$ and $y$ in  $\G$ is a sequence $\Pt=\lp x=v_{1},\ldots,v_{k}=y\rp$ of $k\geq 2$ distinct vertices such that $\{v_{i},v_{i+1}\}\in \E$ for every $i=1,\ldots,k-1$ and we denote by $\Pi_{xy}$ the collection of all paths from $x$ to $y$ in $\G$. We denote by $V(\Pt)\subseteq V$ and $\E(\Pt)\subseteq \E$ the set of vertices and edges of the path $\pi$, respectively. When clear from the context, and to improve the readability of sub- and super-scripts, we will set $P\equiv V(\Pt)$.

We say that the concentration matrix $\Kita$ of $\Xbf_{V}$ \emph{implies} the graph $\G=(V, \E)$ if every nonzero off-diagonal entry of $\Kita$ corresponds to an edge in $\G$. The  \emph{concentration graph model}
\citep{cox1996multivariate}  with graph $\G$ is the family of multivariate normal distributions whose concentration matrix implies $\G$. The latter model has also been called a \emph{covariance selection model} \citep{dempster1972covariance} and a
\emph{graphical Gaussian model}
\citep{whittaker1990graphical}; we refer the reader to \citet{lauritzen1996graphical} for details and discussion on this type of model.
\subsection{Path weights}
Let $V$ be a finite set and $\G=(V, \E)$ be an undirected graph. Furthermore, let $\Pt$ be a path from $x$ to $y$ in $\G$ and $\Gamma\equiv\Gamma_{VV}$ a positive definite matrix indexed by the elements of V. We set
\begin{eqnarray}\label{EQN:path.weight1}
    \w(\Pt, \Gamma)
    &\equiv& (-1)^{|P|+1}\;|\Gamma_{PP}|\;\prod_{\{u,v\}\in \E(\Pt)} \{\Gamma^{-1}\}_{uv}\,,
\end{eqnarray}
where $|P|$ denotes the cardinality of $P=V(\Pt)$ whereas $|\Gamma_{PP}|$ is the determinant of $\Gamma_{PP}$. \citet{jones2005covariance} introduced (\ref{EQN:path.weight1}) in an alternative formulation that relies on the equality
\begin{eqnarray}\label{EQN:path.weight2}
  |\Gamma_{PP}|=\frac{|\Theta_{\bar{P} \bar{P}}|}{|\Theta|}
\end{eqnarray}
where $\Theta=\Gamma^{-1}$, and
with the convention that $|\Theta_{\bar{P} \bar{P}}|=1$ whenever $\bar{P}=\emptyset$.
\begin{thm}[\citet{jones2005covariance}]\label{THM:jones.west}
Let $\Kita=\Sigma^{-1}$ be the concentration matrix of  $\Xbf_{V}$. If $\Kita$ implies the graph $\G=(V, \E)$ then for every $x,y\in V$  it holds that
\begin{eqnarray}\label{EQN:in.thm.jones.west}
\sigma_{xy}
    =\sum_{\Pt\in \Ptxy_{xy}} \w(\Pt, \Sigma)=\sum_{\Pt\in \Ptxy_{xy}}  (-1)^{|P|+1}\;\frac{|\Kita_{\bar{P} \bar{P}}|}{|\Kita|}\;\prod_{\{u,v\}\in \E(\Pt)} \kappa_{uv}.
\end{eqnarray}
\end{thm}
We call $\w(\Pt, \Sigma)$ in (\ref{EQN:in.thm.jones.west}) the \emph{path weight of $\Pt$ relative to $\Xbf_{V}$}. Furthermore, we will refer to (\ref{EQN:in.thm.jones.west}) with the name of the \emph{covariance decomposition over $\G$} because it gives a decomposition of $\sigma_{xy}$ into the sum of the path weights for  all the paths connecting the two vertices in $\G$. We recall that another interesting decomposition of the covariance in Gaussian models in terms of walk-weights can be found in \citet{malioutov2006walk} and references therein. Unlike paths, walks can cross an edge multiple times.
\subsection{The partial vector correlation coefficient}
We denote by $\rho_{xy}$ the \emph{correlation coefficient} of the variables $X_{x}$ and $X_{y}$, with $x,y\in V$. Furthermore, we write  $\rho_{xy\given V\backslash \{x, y\}}$ to denote the \emph{partial correlation coefficient} of  $X_{x}$ and $X_{y}$ given $\Xbf_{V\backslash \{x, y\}}$, and recall that \citep[p.~130]{lauritzen1996graphical}
\begin{eqnarray}\label{EQN:partial.correlation.from.K}
\rho_{xy\given V\backslash \{x, y\}}=\frac{-\kappa_{xy}}{\sqrt{\kappa_{xx}\kappa_{yy}}}\,.
\end{eqnarray}
In the literature, different quantities have been introduced to provide a generalization of the concept of (partial) correlation from pairs of variables to pairs of vectors; see \citet{robert1976unifying}, \citet[][Section~6.5.4]{mardia1979multivariate}, \citet[][p.~485]{timm2002applied}  and \citet[][Section~5.6]{kim2006univariate} for a review of measures of correlation between vectors.
The rest of this section is devoted to a coefficient, called  the \emph{partial vector correlation}, which plays a central role in this paper because it naturally arises in the theory of path weights. As shown below, this coefficient can be obtained as a function of certain canonical correlations and this can be used to assess its connections with other more common measures of association between vectors such as, for instance, the RV-coefficient \citep[see][for details]{robert1976unifying}.

For a pair $A,B\subseteq V$, with $A\cap B=\emptyset$, \citet{hotelling1936relations} introduced the \emph{vector alienation coefficient} defined as
$\lambda_{\vc{A}{B}}\equiv|\Sigma_{A\cup B A\cup B}|/\left(|\Sigma_{AA}|\times|\Sigma_{BB}|\right)$.
Notice that the sampling version of $\lambda_{\vc{A}{B}}$ is the Wilks' lambda, used to test the independence of  $\Xbf_{A}$ and $\Xbf_{B}$ under normality. Furthermore,
\begin{eqnarray}\label{EQN:lambda.and.cancorr}
\lambda_{\vc{A}{B}}= \prod_{i=1}^{r} (1-\varrho_{i}^2)\,,
\end{eqnarray}
where $\varrho_{i}$, for $i=1,\ldots,r$, is the $i$-th canonical correlation between $\Xbf_{A}$ and $\Xbf_{B}$ and $r=\min(|A|, |B|)$; see  \citet[][Section~6.5.4]{mardia1979multivariate} and \citet[][p.~485]{timm2002applied}.

The vector alienation coefficient was used by \citet{rozeboom1965linear} to define the \emph{vector correlation coefficient} given by
\begin{eqnarray}\label{EQN:vector.correlation}
  \rho_{\vc{A}{B}}\equiv\sqrt{1-\lambda_{\vc{A}{B}}}\,,
\end{eqnarray}
and it is easy to check that, for $A=\{x\}$, $\rho^{2}_{\vc{A}{B}}$ coincides with the square of the multiple correlation coefficient so that if also $B=\{y\}$ then $\rho_{\vc{A}{B}}^{2}=\rho_{xy}^{2}$
\citep[see also][p.~485]{timm2002applied}.

Consider a subset $C\subseteq V$ such that $A\cap C=B\cap C=\emptyset$. \citet{rozeboom1965linear} generalized (\ref{EQN:vector.correlation}) to the \emph{partial vector correlation coefficient} as follows,
\begin{eqnarray}\label{EQN:partial.vector.correlation}
\rho_{\cvc{A}{B}{C}}
  =\sqrt{1-\lambda_{\cvc{A}{B}{C}}},
  \qquad\mbox{where}\qquad
  \lambda_{\cvc{A}{B}{C}}
  =\frac{|\Sigma_{A\cup B A\cup B\given C}|}{|\Sigma_{AA\given C}||\Sigma_{BB\given C}|}.
\end{eqnarray}
We remark that the covariance matrices we consider are assumed to be positive definite so that $0\leq \rho_{\cvc{A}{B}{C}}<1$.
Furthermore, $\rho_{\vc{A}{B}}=\rho_{\cvc{A}{B}{\emptyset}}$, and we use the convention that $\rho_{\cvc{A}{B}{C}}=0$ whenever either $A=\emptyset$ or $B=\emptyset$. Note that, for  $A=\{x\}$ and $B=\{y\}$ it holds that $\rho_{\cvc{A}{B}{C}}^{2}=\rho_{xy\given C}^{2}$ that is the square of the partial correlation.
\section{Networked partial covariance and correlation}\label{SEC:NPC}
The decomposition of the covariance $\sigma_{xy}$ over an undirected graph in (\ref{EQN:in.thm.jones.west}) associates a weight to every path $\Pt$ between $x$ and $y$ in $\G$. Hence, the weight $\w(\Pt, \Sigma)$ represents the contribution of the path $\Pt$ to the covariance $\sigma_{xy}$ and from this perspective it is appealing to investigate this quantity as a measure of association between $X_{x}$ and $X_{y}$. However, one cannot readily exploit the covariance decomposition over paths because the interpretation of path weights is unclear and is still an open problem.  More specifically,
it follows from equation (\ref{EQN:partial.correlation.from.K}) that the term $(-1)^{|P|+1}$ in equation $(\ref{EQN:in.thm.jones.west})$ is such that $\w(\Pt, \Sigma)$ has the same sign as the product of the partial correlations corresponding to the edges of the path but, otherwise, it is not clear what the meaning is of the value taken by a path weight. In this section, we address this question by focusing on the special and relevant case of \emph{single-edge paths}, which are paths made of a single edge.

If an edge is missing from the graph $\G$, say $\{x,y\}\not\in \E$, then the corresponding partial covariance is equal to zero, $\sigma_{xy\given V\backslash \{x,y\}}=0$, and for this reason partial covariances and partial correlations are regarded as natural measures to be associated with the edges of the graph. The following theorem shows that the weight $\w(\lp x,y\rp, \Sigma)$ of a single-edge path $\lp x,y\rp$ is a quantity that involves not only the partial covariance associated with the edge, but also a vector correlation coefficient.
\begin{thm}\label{THM:NPCov-from-path-weight}
Let $\Kita=\Sigma^{-1}$ be the concentration matrix of $\Xbf_{V}$. If $\Kita$ implies the graph $\G=(V, \E)$, then for  every $\{x,y\}\in\E$ it holds that
\begin{eqnarray}\label{EQN:net.cov.definition}
  \w(\lp x,y\rp, \Sigma)
  &=& \frac{\sigma_{xy\given V\backslash \{x,y\}}}{1-\rho^{2}_{\vc{xy}{V\backslash \{x,y\}}}},
\end{eqnarray}
where we have used the suppressed notation $\rho_{\vc{xy}{B}}=\rho_{\vc{\{x,y\}}{B}}$.
\end{thm}
\begin{proof}
See Appendix~\ref{APP:00A}
\end{proof}
In what follows, we denote the weight of the single-edge path $\lp x,y\rp$ more compactly as
\begin{eqnarray*}
  \ew_{\ewsub{xy}{V\backslash\{x,y\}}}\equiv\w(\lp x,y\rp, \Sigma)
\end{eqnarray*}
and refer to this quantity as a \emph{networked partial covariance}. When the edge $\{x,y\}$ is missing, $\sigma_{xy\given V\backslash \{x,y\}}=0$ and, therefore, $\ew_{\ewsub{xy}{V\backslash \{x,y\}}}=0$. Moreover, $\ew_{\ewsub{xy}{V\backslash \{x,y\}}}$ and $\sigma_{xy\given V\backslash \{x,y\}}$ have the same sign, and $|\ew_{\ewsub{xy}{V\backslash \{x,y\}}}|\geq |\sigma_{xy\given V\backslash \{x,y\}}|$.

Furthermore, the ratio in equation (\ref{EQN:net.cov.definition}) provides a clear interpretation of the edge weight $\ew_{\ewsub{xy}{V\backslash\{x,y\}}}$ because it shows that $\ew_{\ewsub{xy}{V\backslash\{x,y\}}}$ is obtained by combining the information that is provided by $\sigma_{xy\given V\backslash \{x,y\}}$ and $\rho_{\vc{xy}{V\backslash \{x,y\}}}$. More concretely, the networked partial covariance is computed by multiplying the partial covariance $\sigma_{xy\given V\backslash \{x,y\}}$  by $1/\{1-\rho_{\vc{xy}{V\backslash \{x,y\}}}\}$, which is always greater than or equal to 1 and an increasing function of $\rho_{\vc{xy}{V\backslash \{x,y\}}}$. Furthermore, it is worth noticing that $\sigma_{xy\given V\backslash \{x,y\}}$  and $\rho_{\vc{xy}{V\backslash \{x,y\}}}$ provide two distinct pieces of information.
\begin{enumerate}
  \item[(a)] The information provided by $\sigma_{xy\given V\backslash \{x,y\}}$ concerns the presence of the edge  $\{x,y\}$ in $\G=(V, \E)$ because $\sigma_{xy\given V\backslash \{x,y\}}\neq 0$ implies $\{x,y\}\in \E$. More concretely, it equals the covariance of $X_{x}$ and $X_{y}$ computed after the two variables have been linearly adjusted for the remaining variables in the network. Hence, $\sigma_{xy\given V\backslash \{x,y\}}$ provides no information on the strength of the linear association between $X_{x}$ and $X_{y}$ and the remaining variables in the network. In other words, $\sigma_{xy\given V\backslash \{x,y\}}$ can be regarded as an \lq outer\rq{} measure of the association encoded by the edge $\{x,y\}$, because the way in which $\{x,y\}$ is connected with the rest of the network, plays no role in its computation. This kind of interpretation is even stronger in the case where the variables are jointly Gaussian, because in this case  $\sigma_{xy\given V\backslash \{x,y\}}$ is the covariance of the conditional distribution of $\Xbf_{\{x,y\}}|\Xbf_{V\backslash\{x,y\}}$.
  \item[(b)] The vector correlation $\rho_{\vc{xy}{V\backslash \{x,y\}}}$ is a measure of the strength of the association between  $\Xbf_{\{x,y\}}$ and the remaining variables $\Xbf_{V\backslash\{x,y\}}$, and provides no information on whether $x$ and $y$ are joined by an edge. Regardless of whether $\{x,y\}$ is an edge of the graph, when the pair $\{x, y\}$ is disconnected from the rest of the network then $\rho_{\vc{xy}{V\backslash \{x,y\}}}=0$ and, consequently, $\ew_{\ewsub{xy}{V\backslash \{x,y\}}}=\sigma_{xy\given V\backslash \{x,y\}}$.
\end{enumerate}
In summary, the weight $\ew_{\ewsub{xy}{V\backslash\{x,y\}}}$ synthesizes in a single quantity the strength of the partial covariance between $X_{x}$ and $X_{y}$, and the strength of the vector correlation between $X_{\{x,y\}}$ and the remaining variables in the network. This interpretation motivates the name of \emph{networked partial covariance}.

Just as covariances need to be normalized into correlations to enable their comparison, we provide also the normalized version of equation (\ref{EQN:net.cov.definition}) that we shall call the \emph{networked partial correlation}:
\begin{eqnarray}\label{EQN:w.ast.definition}
  \ewnorm_{\ewsub{xy}{V\backslash \{x,y\}}}
  \equiv\frac{\ew_{\ewsub{xy}{V\backslash \{x,y\}}}}{\sqrt{\sigma_{xx\given V\backslash \{x,y\}}\,\sigma_{yy\given V\backslash \{x,y\}}}}
  = \frac{\rho_{xy\given V\backslash \{x,y\}}}{1-\rho^{2}_{\vc{xy}{V\backslash \{x,y\}}}}\,.
\end{eqnarray}
Although expression (\ref{EQN:w.ast.definition}) is a normalized quantity and, therefore, comparable between edges from the same graph, it may take values outside the interval $[-1, 1]$. The following example gives a simplified setting that makes it clear how the networked partial correlation can be regarded as an \lq inflated\rq{} version of the partial correlation to keep into account how the edge is embedded in the network.
\begin{exmp}\label{EXA:equal.pc}
Consider the case where $|V|=9$ and the concentration matrix $\Kita$ of $\Xbf_{V}$ induces the graph $\G=(V, \E)$ in Fig.~\ref{FIG:two.graphs.examples}.
\begin{figure}
 \centering
   {\includegraphics[scale=0.8]{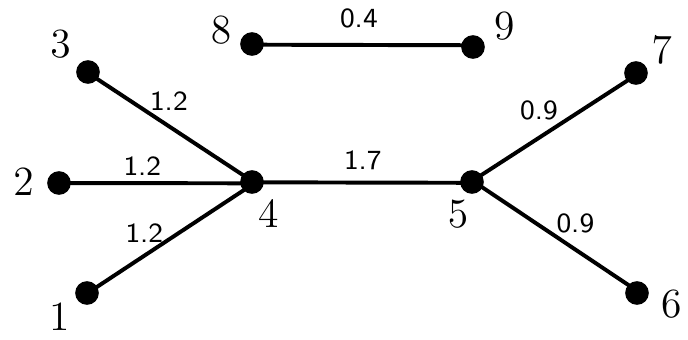}}
\caption{Undirected graph of the Example~\ref{EXA:equal.pc}. Values on the edges correspond to networked partial correlations.}
 \label{FIG:two.graphs.examples}
\end{figure}
More specifically,
we take $\Kita$ to have unit diagonal and off-diagonal elements $\kappa_{uv}=-0.4$ for every $\{u,v\}\in \E$ and $\kappa_{uv}=0$ otherwise. The simplified structure of the  concentration matrix in this example makes it easy to appreciate the differences existing between partial correlations and networked partial correlations. Indeed, in this case, the partial correlations $\rho_{uv\given V\backslash\{u,v\}}$ for $\{u,v\}\in \E$ put all the edges of the graph on an equal footing because they are all equal to $0.4$.
However, the networked partial correlations, whose values are reported in Fig.~\ref{FIG:two.graphs.examples}, are not constant and, in this case, their differences depend only on the structure of the graph.
Indeed,
the edge $\{8, 9\}$ is disconnected from the rest of the vertices so  that the values of its networked partial correlation and partial correlation coincide; i.e. $\ewnorm_{\ewsub{89}{V\backslash \{8,9\}}}=\rho_{89\given V\backslash\{8,9\}}=0.4$. The edge $\{4, 5\}$ has the largest number of connections with other vertices in the graph and, accordingly, its networked partial correlation takes the largest value $\ewnorm_{\ewsub{45}{V\backslash \{4,5\}}}=1.7$. More generally, in this example, the value of the networked partial correlation of every edge is proportional to the number of vertices adjacent to the edge.
\end{exmp}
\section{Limited-order networked partial covariance decomposition}\label{SEC:limited.order.NPC.dec}
In practical applications, it is common to deal with \emph{limited-order} partial covariances, which are partial covariances $\sigma_{xy\given Q}$ with $Q\cup \{x, y\}\subset V$, rather than with
\emph{full-order} partial covariances $\sigma_{xy\given V\backslash \{x,y\}}$. Typically, this is due to the presence of unobserved variables, possibly not explicitly considered in the analysis or because the number of variables exceeds the sample size, so that the sample covariance matrix has not full rank, thereby making the computation of full-order partial covariances unfeasible; see  \citet{castelo2006robust}, \citet{zuo2014biological} and references therein.
In these cases, it is therefore also sensible to work with limited-order path weights rather than full-order path weights. Consider the concentration matrix $\Kita=\Sigma^{-1}$ of $\Xbf_{V}$ that implies the graph $\G=(V, \E)$.
For a subset $Q\subset V\backslash \{x, y\}$ we define the limited-order weight of the single-edge path $\lp x,y\rp$ as the weight of $\lp x,y\rp$ relative to $\Xbf_{Q\cup \{x,y\}}$; formally
$\ew_{\ewsub{xy}{Q}}\equiv\w(\lp x,y\rp, \Sigma_{Q\cup\{x,y\}Q\cup\{x,y\}})$.
Since it follows from Theorem~\ref{THM:NPCov-from-path-weight} that
\begin{eqnarray}\label{EQN:l.o.NPC.definition}
  \ew_{\ewsub{xy}{Q}}
  &=& \frac{\sigma_{xy\given Q}}{1-\rho^{2}_{\vc{xy}{Q}}},
\end{eqnarray}
we refer to $\ew_{\ewsub{xy}{Q}}$ as a \emph{limited-order networked partial covariance}. For $Q=\emptyset$ it holds that $\ew_{\ewsub{xy}{Q}}=\sigma_{xy.Q}=\sigma_{xy}$, and therefore, the (marginal) covariance can be regarded as a special case of both the limited-order partial covariance and the  limited-order networked partial covariance. It follows from (\ref{EQN:l.o.NPC.definition}) that the \emph{limited-order networked partial correlation} can be defined as
\begin{eqnarray*}
  \ewnorm_{\ewsub{xy}{Q}}
  =\frac{\ew_{\ewsub{xy}{Q}}}{\sqrt{\sigma_{xx\given Q}\,\sigma_{yy\given Q}}}
  = \frac{\rho_{xy\given Q}}{1-\rho^{2}_{\vc{xy}{Q}}}.
\end{eqnarray*}

To interpret the meaning of any limited-order quantity properly it is necessary to clarify how such quantity is affected by the marginalization over the variables that are excluded from the analysis. More specifically, when the relevant limited-order quantity  is used to describe the association that is represented by an edge of the graph, then it is of interest to investigate what the role that is played by the structure of the full, unobserved, network $\G_{V}$ is in the specification of such a quantity. In the following theorem we give a rule to decompose a limited-order networked partial covariance $\ew_{\ewsub{xy}{Q}}$ over the paths between $x$ and $y$ in $\G_{V\backslash Q}$. This clarifies what the information that is provided by a networked partial covariance is, thereby providing a theoretical justification for its use.
\begin{thm}\label{THM:path.update}
Let $\Kita=\Sigma^{-1}$ be the concentration matrix of  $\Xbf_{V}$. If $\Kita$ implies the graph $\G=(V, \E)$ then for every $x,y\in V$ and $Q\subseteq V\backslash \{x, y\}$ it holds that
\begin{eqnarray}\label{EQN:path.relation}
\ew_{\ewsub{xy}{Q}}
&=&  \sum_{\Pt\in \Ptxy_{xy}; V(\Pt)\subseteq V\backslash Q}
\w(\Pt, \Sigma)
\times
(1-\rho^{2}_{\cvc{P\backslash \{x, y\}}{Q}{\{x,y\}}})
\end{eqnarray}
where $P=V(\Pt)$.
\end{thm}
\begin{proof}
See Appendix~\ref{APP:00A}
\end{proof}
First, when $Q=\emptyset$ equation (\ref{EQN:path.relation}) coincides with (\ref{EQN:in.thm.jones.west}) so that the limited-order partial covariance decomposition in Theorem~\ref{THM:path.update} includes, as a special case, the covariance decomposition of \citet{jones2005covariance}, given in  Theorem~\ref{THM:jones.west}. Second, for $Q=V\backslash \{x,y\}$ equation (\ref{EQN:path.relation}) simplifies to  $\ew_{\ewsub{xy}{V\backslash\{x,y\}}}=\w(\lp x,y\rp, \Sigma)$.
More generally, the decomposition of the limited-order networked partial covariance given in Theorem~\ref{THM:path.update} enables us to understand the connection between the weight of a path in a graph derived from a multivariate distribution, and the weight of a path in a graph derived from a marginal distribution. Concretely, it shows that every path $\Pt\in \Pi_{xy}$ such that $V(\Pt)\cap Q=\emptyset$ contributes to the value of $\ew_{\ewsub{xy}{Q}}$ with the proportion $1-\rho^{2}_{\cvc{P\backslash\{x,y\}}{Q}{\{x,y\}}}$ of its weight $\w(\Pt, \Sigma)$. More importantly, a path between two vertices $x$ and $y$ contributes to the value of $\ew_{\ewsub{xy}{Q}}$ only if all its vertices, except for $x$ and $y$, have been marginalized over. This means that any path with at least one endpoint not equal to $x$ or $y$, and any path between $x$ and $y$ involving at least one vertex in $Q$, plays no role in the computation of $\ew_{\ewsub{xy}{Q}}$. To make the rules for limited-order networked partial covariance decomposition more concrete,  Appendix~\ref{APP:00C}  gives a detailed description of the case where $|V|=4$ and $|Q|=1$.

Because any networked partial covariance is a path weight, an appealing feature of equation (\ref{EQN:path.relation}) is that both the term $\ew_{\ewsub{xy}{Q}}$ in the left hand side and the terms $\w(\Pt, \Sigma)$ in the right hand side are path weights. This confers consistency to equation (\ref{EQN:path.relation}) that can thus be regarded as a rule to update the weight of single-edge paths when the multivariate system is marginalized over some variables.  This motivates the use of the networked partial covariance as a natural generalization of the covariance. From this viewpoint, it is also worth noting that, by multiplying the left- and right-hand side of (\ref{EQN:path.relation}) by $1-\rho^{2}_{\vc{xy}{Q}}$, Theorem~\ref{THM:path.update} can be restated to provide a rule to decompose  $\sigma_{xy\given Q}$. However, consistency of interpretation between the left- and the right-side of the equation is lost in this case.

\section{Analysis of genetic interactions in yeast}\label{SEC:genetic.interactions}

\subsection{Data preparation and estimation methods}\label{SEC:data.preparation.est.methods}
\citet{costanzo2010genetic} generated quantitative genetic interaction profiles
in a systematic way for about 75\% of all the genes in yeast, using a technique called
synthetic genetic array (SGA) analysis. This technique enabled the quantification
for 6,647,235 gene pairs in yeast of the fitness effect of a double mutant with
respect to the expected effect calculated from the combination of two single mutants.
This quantification was provided through the so-called SGA scores that also have an
associated $p$-value that captures how reliable they are \citep{baryshnikova2010quantitative}.
This reliability is measured through a combination of the observed variation across
four experimental replicates, with estimates of the background log-normal error
distributions for the corresponding mutants \citep{baryshnikova2010quantitative,costanzo2010genetic}.
We downloaded those SGA scores and $p$-values and filtered them to discard pairs displaying
a defective experimental procedure, such as a missing SGA score, or duplicated gene pairs
with SGA scores of opposite sign. Between two SGA scores of the same sign produced by a
duplicated gene pair, we kept the SGA score with lowest $p$-value as suggested in
\citep{costanzo2010genetic}. After this filtering step, we kept 5,195,591 gene pairs
involving 4457 genes. We used these 5 million SGA scores as gold-standard for the
fitness effect of genetic interactions in yeast (see Supplementary Materials).

To demonstrate the usefulness of networked partial correlations in this context,
we used gene expression data produced by \citet{Brem2005landscape} from a cross
between two yeast strains: a wild-type (RM11-1a) and a lab strain (BY4716). These
two strains were crossed by \citet{Brem2005landscape} to generate $n=112$
segregants whose gene expression was profiled with microarray chips. We downloaded
and processed the resulting raw data as described in \citet{tur2014mapping}
leading to a normalized gene expression data matrix formed by $p=6216$ genes and
$n=112$ samples.

The calculation of the networked partial correlation from expression data between
two given genes involves the estimation of two quantities
(see equation~(\ref{EQN:w.ast.definition})): (i) the partial correlation between these
two genes; and (ii) the vector correlation between this pair of genes and the
rest of the genes. Because the number of genes, $p$, is much larger than the
number of samples, $n$, i.e., $p\gg n$, the calculation of these two quantities
is not straightforward and requires the use of statistical methods specifically
tailored to deal with high-dimensional data where $p\gg n$.

\begin{figure}[p]
\centerline{\includegraphics[width=0.8\textwidth]{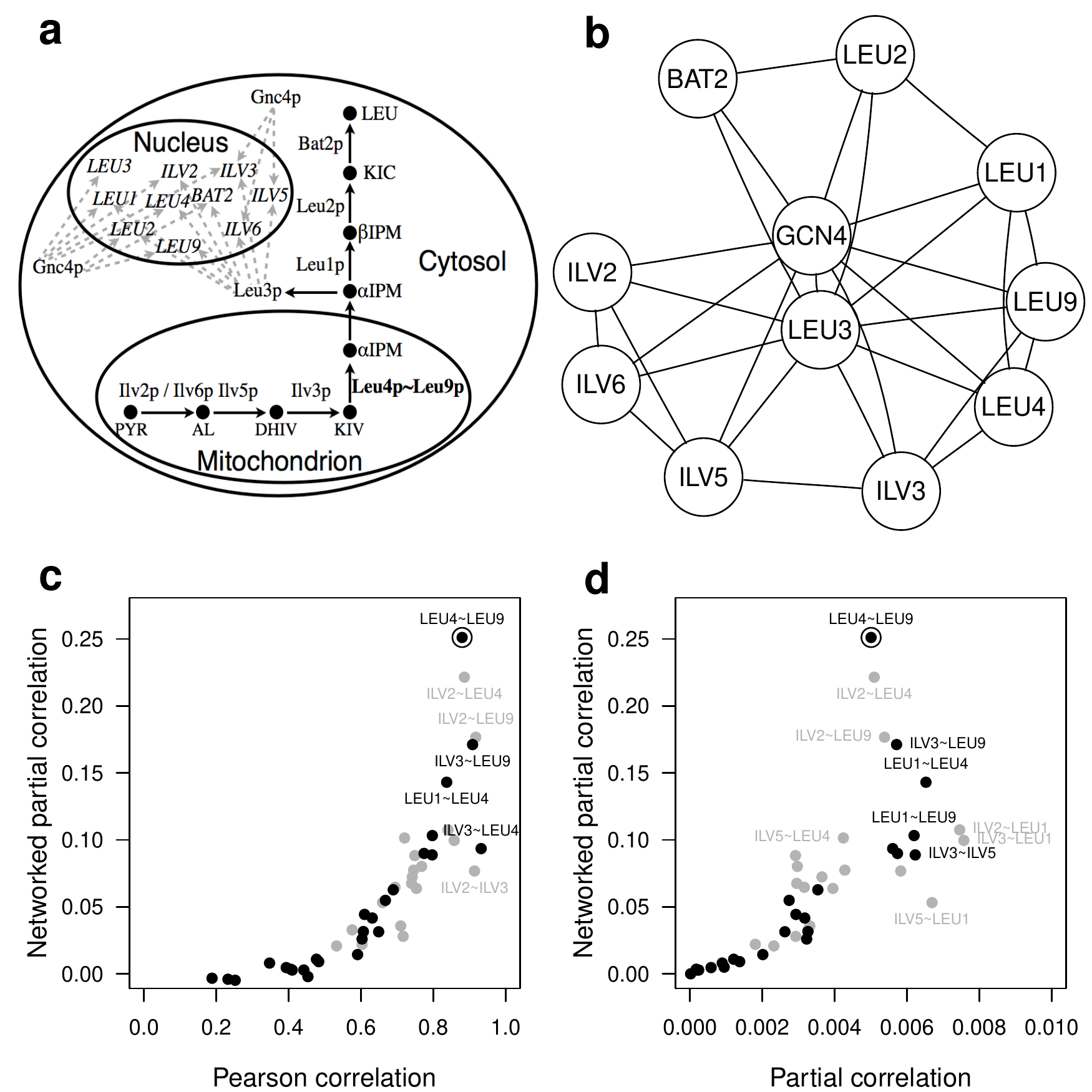}}
\caption{Analysis of the leucine biosynthesis pathway. (a) A schematic
representation of the pathway \citep[see][]{kohlhaw2003leucine,chin2008dynamics}.
Solid dots and arrows indicate metabolites and metabolic reactions. Dashed arrows
indicate transcriptional regulatory associations. Metabolite abbreviations are placed
next to the corresponding metabolite. Enzyme protein names are placed next
to the corresponding metabolic reaction. Slash {\tiny /} and tilde {\tiny$\sim$} symbols
indicate protein complex and genetic interactions, respectively. (b) Undirected
graph representing direct functional associations between genes involved in the
pathway depicted in (a). (c) Networked partial correlation values on the
$y$-axis as a function of Pearson correlation values on the $x$-axis. Black and
grey dots correspond to present and missing edges in (b), respectively. A circle
indicates the only known genetic interaction among genes in (b). (d) The same as (c)
but the values on the $x$-axis are partial correlations.}
\label{fig:PACvsNPCleupwy}
\end{figure}

We estimated partial correlation coefficients and their $p$-values for the null
hypothesis of zero-partial correlation by using the empirical Bayes
method of \citet{schafer2005empirical} that works by calculating a shrinkage
estimate of the inverse covariance and is implemented in the R package GeneNet.
To estimate vector correlations we exploited expression (\ref{EQN:lambda.and.cancorr})
and used the sparse canonical correlation analysis technique of
\citet{witten2009penalized} implemented in the R package PMA. Full details
on how the data analysis was conducted are available in the Supplementary
Materials. Data and source code of the R scripts reproducing the results in this
section are available at \url{http://functionalgenomics.upf.edu/supplements/NPC4GI}.

\subsection{Analysis of the leucine biosynthesis pathway}

The gene expression data by \citet{Brem2005landscape} were generated by first
crossing two different strains of yeast, one of them containing the deletion
of the \textit{LEU2} gene that participates in the leucine biosynthesis pathway.
Then, gene expression was profiled in the resulting collection of $n=112$
segregants. Because some of these offspring inherited the deletion of the
\textit{LEU2} gene, these gene expression data show a large degree of variability
of expression in genes involved in the leucine biosynthesis pathway, providing the
opportunity to study gene expression changes associated with the activity of this pathway.

The leucine biosynthesis pathway, which is shown in Fig.~\ref{fig:PACvsNPCleupwy}(a),
consists of a number of sequential reactions catalysed by different enzymes that
allow yeast to convert pyruvate (PYR) into leucine (LEU). Among these reactions,
a key role is played by a metabolic intermediate called $\alpha$-isopropylmalate
($\alpha$IPM), which binds to the homodimeric DNA binding protein Leu3p, which is a
transcription factor regulating the expression of all the genes within the pathway.
The transcriptional activity of all genes in the pathway, including \textit{LEU3}
itself, is also regulated by the transcription factor Gcn4p. See
\citep{kohlhaw2003leucine,chin2008dynamics} for a more comprehensive description of
this pathway.

$\alpha$IPM is synthesised by either of the two enzymes encoded by the genes
\textit{LEU4} and \textit{LEU9} \citep{kohlhaw2003leucine}, who are paralogues
and form a duplicated gene pair that arose from the whole genome duplication
of yeast. It is well known that the deletion of only one of these
two genes is not sufficient to create a leucine-auxotrophic yeast mutant that
would require a supply of leucine for growth \citep{kohlhaw2003leucine}.
Consistent with this observation, the gene pair \textit{LEU4}-\textit{LEU9}
forms a genetic interaction whose double mutation produces a fitness defect
that is more severe than what is expected from the combination of the single
mutants \citep{deluna2008exposing}. All other possible interactions between genes
that are involved in the pathway (Fig.~\ref{fig:PACvsNPCleupwy}b) were either absent
from the catalogue of quantitative genetic interaction profiles analysed in this
paper \citep{costanzo2010genetic} or did not have a negative and significant
(false discovery rate (FDR) $<1$\%) SGA interaction score.

One of the simpler buffering relationships behind a genetic interaction is
the positive coexpression of two genes and, accordingly, we analysed only those
pairs of genes in this pathway with positive Pearson, partial and networked
partial, correlations, previously calculated from the expression data.

The comparison between these quantities shown in Figs.~\ref{fig:PACvsNPCleupwy}c
and \ref{fig:PACvsNPCleupwy}d reveals that the only known genetic interaction
\textit{LEU4}-\textit{LEU9} has the largest networked partial correlation among
all the gene pairs, which is not so for Pearson or partial correlations. The
following three gene pairs ranked by the networked partial
correlation, \textit{ILV2}-\textit{LEU4}, \textit{ILV2}-\textit{LEU9} and
\textit{ILV3}-\textit{LEU9}, involve each of the two genes forming the
known \textit{LEU4}-\textit{LEU9} genetic interaction and the other intervening
genes \textit{ILV2} and \textit{ILV3} are upstream of $\alpha$IPM, where they
have more chance to affect its synthesis and, therefore, the entire operation
of the pathway \citep{chin2008dynamics}.

\begin{figure}[ht]
\centerline{\includegraphics[width=1.0\textwidth]{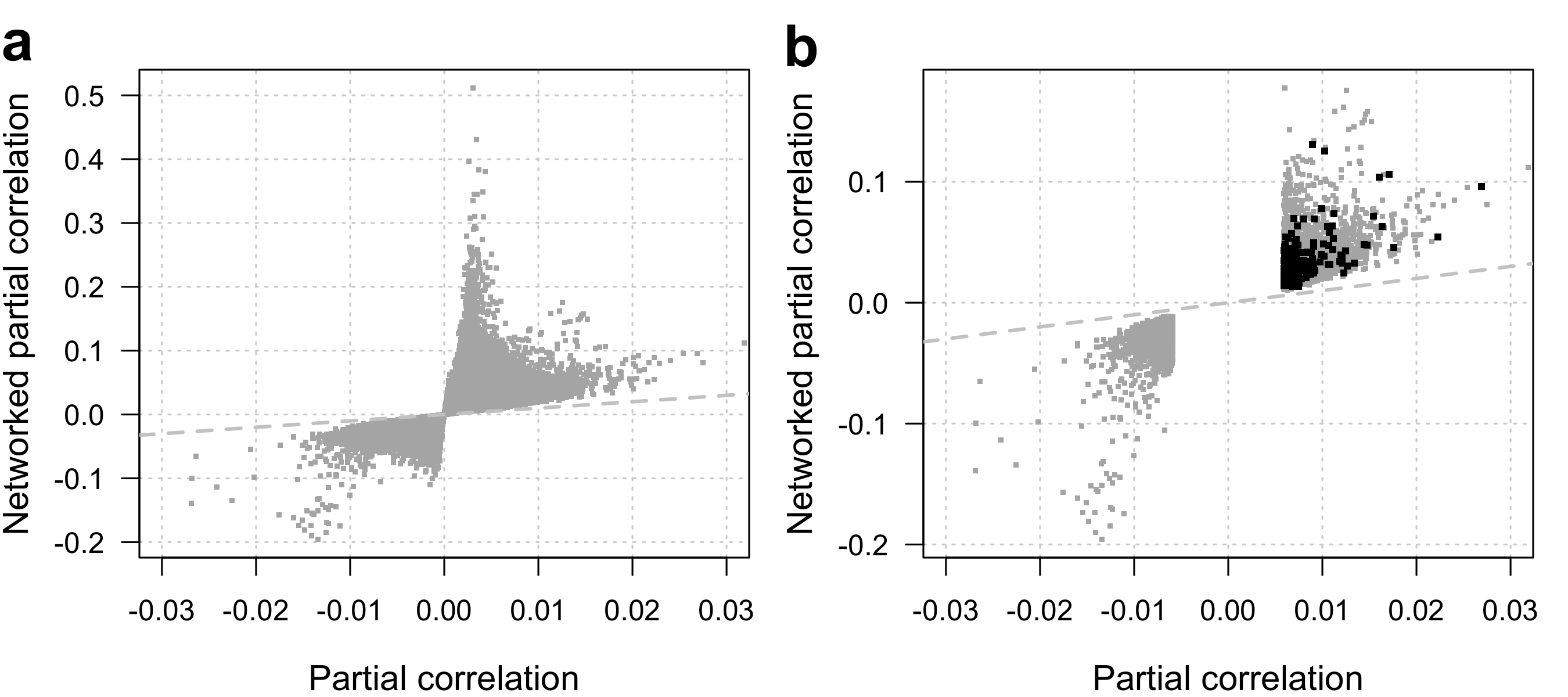}}
\caption{Networked partial correlations on the $y$-axis as a function of the
         corresponding partial correlations on the $x$-axis, calculated from
         yeast expression data \citep{Brem2005landscape}. (a) Values for all
         pairs of genes in the expression data set. (b) Values for those
         pairs with significant partial correlations (FDR $< 1$\%). Values for
         the filtered pairs shown in Figures~\ref{fig:SGAcomp} and
         \ref{fig:SGAcomplog} are highlighted in black. The grey dashed line indicates
         the axis where $x=y$ and is provided only as a visual guide.}
\label{fig:PCORvsNPC}
\end{figure}

\subsection{Analysis of quantitative genetic interaction profiles}\label{SEC:identification.of.gen.inter}
In this subsection we analyze the genomewide quantitative interaction profiles
from \citet{costanzo2010genetic}, defined by SGA scores and $p$-values associated
with the profiled gene pairs. There were 4099 genes in common between the 4457 genes
forming pairs with SGA scores and $p$-values, and the 6216 genes with expression
data. We restricted the rest of the analysis to the 3,966,346 pairs formed by
these 4099 genes. A comparison of the values of partial and networked partial,
correlations, shown in Figure~\ref{fig:PCORvsNPC}(a), reveals that differences
between these two quantities grow proportionally to their absolute value. Note
that small values of partial correlation may still become large networked partial
correlation values.

\begin{figure}[ht]
\centerline{\includegraphics[width=0.75\textwidth]{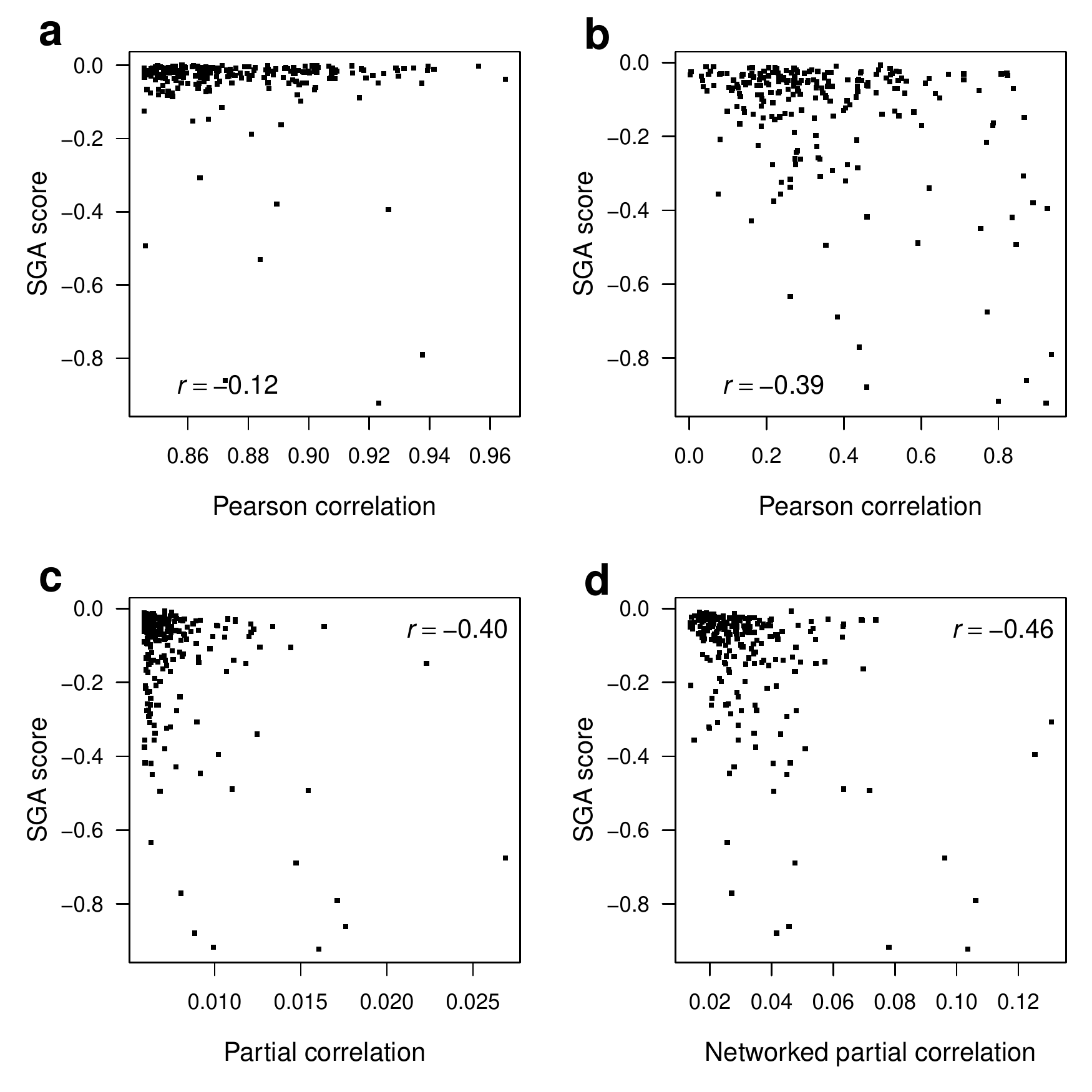}}
\caption{SGA scores as function of the three different gene correlation measures.
         (a) SGA scores on the $y$-axis as function of Pearson correlations on the
         $x$-axis for gene pairs with largest Pearson correlation. There are as
         many pairs as gene pairs with significant partial correlation (FDR $< 1$\%).
         (b) The same as (a) but gene pairs were selected with significant partial
         correlation (FDR $< 1$\%). (c,d) SGA scores on the $y$-axis as function of
         partial and networked partial, correlations, respectively, on the $x$-axis.
         Gene pairs were selected as in (b).}
\label{fig:SGAcomp}
\end{figure}

Positive and negative SGA scores have a very different interpretation. While
negative SGA scores indicate a fitness defect that is more severe than expected,
positive ones identify double mutants whose fitness defect is less severe than
expected \citep{costanzo2010genetic}. For this reason, and to provide a
meaningful comparison between SGA scores and correlation measures, we restricted
the subset of analyzed gene pairs as follows. First, we considered only 87,471
gene pairs with negative and significant SGA scores whose FDR $< 1$\% on the
corrected SGA $p$-value. Second, we further restricted the analysis to gene pairs
showing positive and significant coexpression.

When we considered Pearson correlation coefficients with a corrected $p$-value
of FDR$< 1$\% to define such gene pairs, 6,889 of them were selected. The
association between their SGA scores and their magnitude of the
Pearson correlation was negligible possibly due to the large number of significant
spurious associations (see Supplementary Materials). In contrast, when we
considered significant partial correlation coefficients with FDR$< 1$\%, only
227 gene pairs were selected. To enable a more direct comparison of the
performance of Pearson correlation coefficients we considered also selecting the
top-227 gene pairs with largest positive Pearson correlation values. Selecting
a top number of gene pairs with the largest marginal correlation, such as Pearson
or Spearman, is a common strategy used in computational pipelines for selecting
coexpressed genes potentially forming a genetic interaction
\citep[e.g., ][]{jerby2014predicting}.

\begin{figure}[ht]
\centerline{\includegraphics[width=0.75\textwidth]{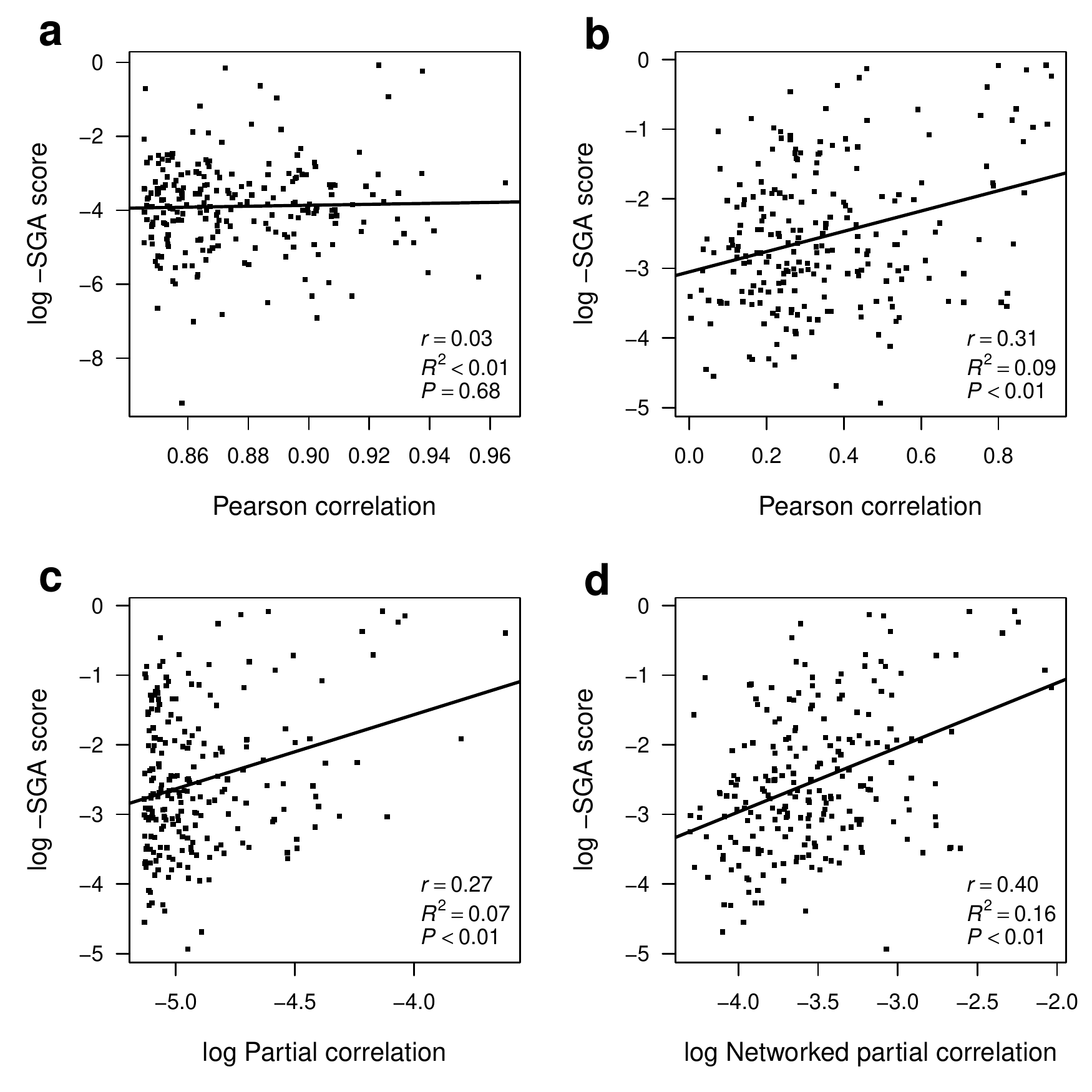}}
\caption{SGA scores as function of different gene correlation measures, on a
         logarithmic scale. Panels (b), (c) and (d) show gene pairs with positive
         and significant partial correlation values at FDR$< 1$\%. Panel (a)
         shows the same number of gene pairs but selected with the largest
         positive Pearson correlation coefficient values.}
\label{fig:SGAcomplog}
\end{figure}

The association of the largest values of Pearson correlation with SGA scores,
remains non-significant, however, as shown in Fig.~\ref{fig:SGAcomp}(a). This
association is greatly improved by using Pearson correlation coefficients only
on gene pairs whose partial correlation is significantly different from zero, as
shown in Fig.~\ref{fig:SGAcomp}(b). Yet, Figs.~\ref{fig:SGAcomp}(c) and \ref{fig:SGAcomp}(d)
show that the association with SGA scores can still improve when partial and
networked partial correlations are used instead on those gene pairs.

Fig.~\ref{fig:SGAcomp} also shows that while larger coexpression values are
associated with larger negative SGA scores, the trend is non-linear. Such a
non-linearity probably arises from the restriction of SGA scores to negative values
and gene coexpression to positive ones, so that a large fraction of pairs accumulate
in values close to zero of both quantities.

To have a clearer picture of the differences between these three coexpression
measures in relationship with SGA scores, we show in Fig.~\ref{fig:SGAcomplog}
the same values in logarithmic scale for absolute SGA scores, partial
correlations and networked partial correlations. These plots reveal that there
is a significant linear relationship between each of these three coexpression
measures and SGA scores, albeit only when gene pairs are selected on the
basis of a test for a zero-partial correlation coefficient; see
Figs.~\ref{fig:SGAcomplog}(b), \ref{fig:SGAcomplog}(c) and \ref{fig:SGAcomplog}(d).
However, among these significant associations, networked partial correlations
explain a larger fraction of the variability of SGA scores ($R^{2}=0.16$) than
Pearson ($R^{2}=0.09$) and partial correlations ($R^{2}=0.07$).

\begin{figure}[ht]
  \centerline{\includegraphics[width=0.75\textwidth]{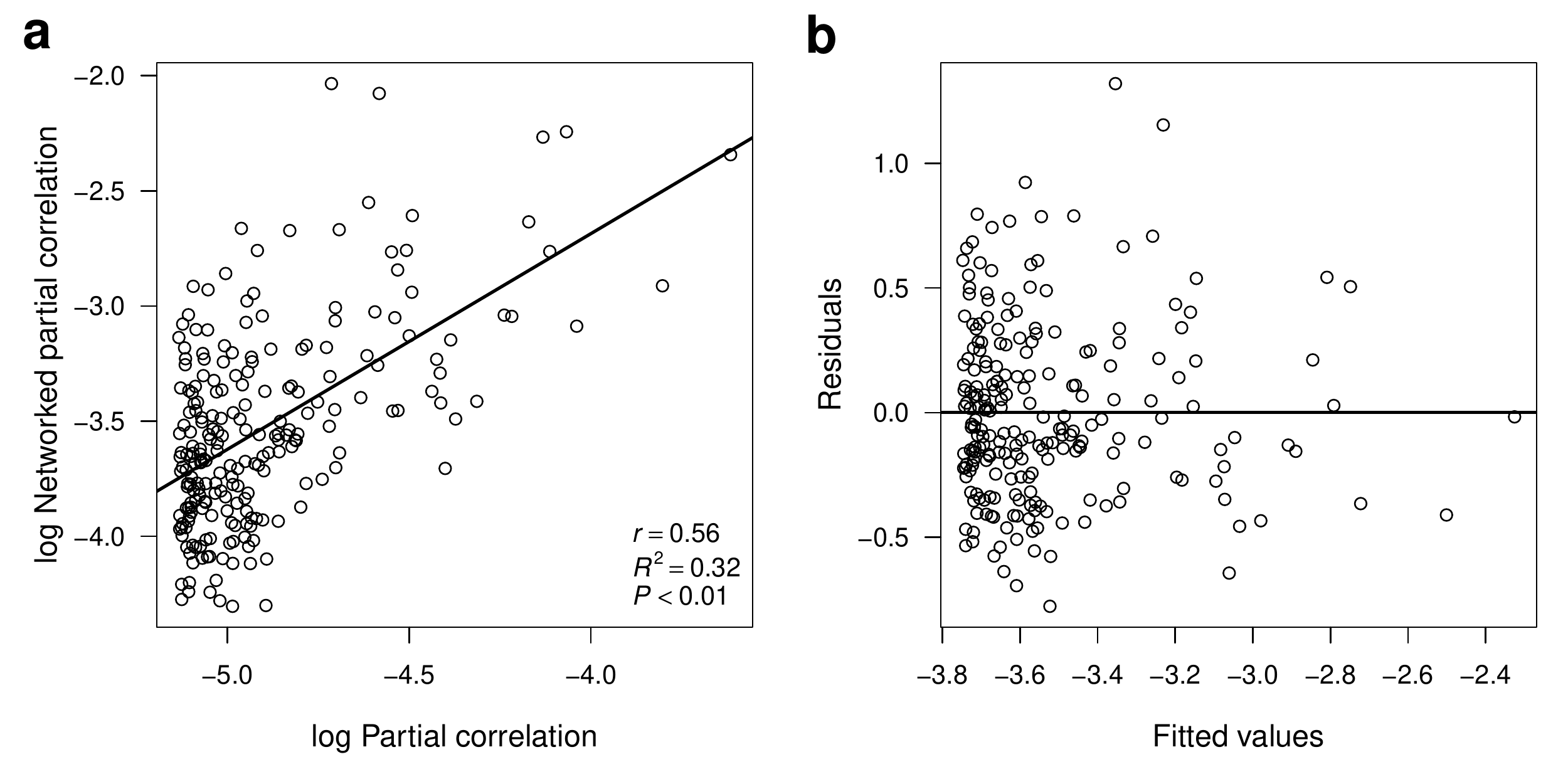}}
  \caption{Regression of networked partial correlations on partial correlations, both on
           a logarithmic scale. (a) Scatter plot of both quantities with the regression line.
           (b) Scatter plot of the residuals of the regression as function of the fitted values.}
  \label{figNPConPAC}
\end{figure}

We also investigated the extent to which networked partial correlations provide
additional information over partial correlations. We first regressed networked
partial correlations on partial correlations, obtaining a significant fit as
expected (Fig.~\ref{figNPConPAC}a). Then, we considered the following three linear
models of the SGA scores: a first model where SGA scores are a linear function of
partial correlation values only, a second model including the residuals of the former
regression (Fig.~\ref{figNPConPAC}b) as an additional term, and a third model as
a linear function of the networked partial correlation values only.

\begin{table}
\begin{center}
\begin{tabular}{lrlrlrl}
\hline
 & Model 1 & & Model 2 & & Model 3 &\\
\hline
(Intercept) &  2.70  & \!\!\!\!\!$^{*}$   &  2.70  & \!\!\!\!\!$^{*}$   & 1.82   & \\
PAC         &  1.07  & \!\!\!\!\!$^{***}$ &  1.07  & \!\!\!\!\!$^{***}$ & 0.29   & \\
NPCresid    &        &                    &  0.83  & \!\!\!\!\!$^{***}$ &        & \\
NPC         &        &                    &        &                    & 0.83   & \!\!\!\!\!$^{***}$ \\
\hline
R$^2$       & 0.07   &                    & 0.16   &                    & 0.16   & \\
RSS         & 213.57 &                    & 193.94 &                    & 193.94 & \\
\hline
\multicolumn{5}{l}{$n=227$, Df $=1$, $F=22.67^{***}$} \\
\hline
\multicolumn{5}{l}{\scriptsize{$^{***}p<0.001$, $^{**}p<0.01$, $^*p<0.05$}}
\end{tabular}
\caption{Comparison of three linear models of the SGA scores as function of
         the partial correlation (PAC) values only (model 1), of the PAC-values and
         the residuals of the networked partial correlation (NPC) values regressed
         on the PAC-values (model 2), and of the PAC-values and the NPC-values (model 3).
         The bottom line gives the sample size $n$, degrees of freedom (Df) and
         $F$-statistic for the analysis of variance of model 1 against model 2 and
         model 3. Both comparisons give exactly the same result.}
\label{tabSGAPACNPCresid}
\end{center}
\end{table}

The results, summarised in Table~\ref{tabSGAPACNPCresid}, show that the two models
including networked partial correlations, or the residuals of their regression on
partial correlations, provide a significantly better fit to SGA scores than
the model that includes partial correlations alone ($p < 0.001$).

\section{Discussion}\label{SEC:discussion}
The theory that was developed by \citet{jones2005covariance} associates a weight
to every path of an undirected graph, and the simple observation that every edge
of the graph is also a path allowed us to introduce the networking partial
covariance, as a novel measure of association between pairs of variables. The
theory of Section~\ref{SEC:limited.order.NPC.dec} shows that, in a context where
the association structure between variables is represented by a network, the
networked partial covariance can be regarded as a natural generalization of the
covariance, thereby providing an additional motivation for its use.

The networked partial covariance can be normalized to obtain a networked partial
correlation. We have shown that the latter has the form of an inflated version of
the partial correlation and that it should be preferred to the partial correlation
to address questions where the relevance the association between two variables also
depends on the strength of the association of the corresponding edge with the rest
of the network. This is so, for instance, for genetic interactions that confer
robustness on cells in response to genetic perturbations. Our analysis of
quantitative genetic interaction profiles in yeast highlights the relevance and
usefulness of the networked partial correlation in this context.

Despite the improved performance of networked partial correlations, the fraction
of variability they explain in quantitative genetic interaction profiles is rather
modest ($R^2=0.16$). However, one should consider the fact that the identification
of genetic interactions on the basis of gene expression data is a very challenging
problem because, on the one hand, buffering relationships are only one of the many
biological mechanisms affecting the expression levels of the genes. On the other
hand, changes in gene expression may occur as a result of multiple types of effects
other than genetic effecst, such as molecular, environmental and technical effects
produced by the profiling instruments. For this reason, the prediction of genetic
interactions is typically based on  multiple biological features
\citep{wong2004combining,zhong2006genome,conde2009human,deshpande2013comparative,jerby2014predicting},
and the information provided by gene correlation measures is only one of the
potential predictors. In this sense, the assessment of the improvement provided by
the introduction of networked partial correlations within current computational
pipelines for the prediction of genetic interactions is of potential interest.

We have estimated networked partial correlations by computing separately the
partial correlation and the vector correlation by means of existing procedures
developed to deal with the case $p\gg n$. More efficient estimates might be
obtained by following a unitary approach to the estimate of this quantity, and
future research should tackle this problem.

\section*{Acknowledgments}
We acknowledge the support of the Spanish Ministry of Economy and Competitiveness
(MINECO) [TIN2015-71079-P], the Catalan Agency for Management of University and
Research Grants (AGAUR) [SGR14-1121] and the European Cooperation in Science and
Technology (COST) [CA15109]. We thank the Joint Editor, the Associate Editor and
the two referees for their comments that have helped to improve the manuscript.

\appendix

\section{Proofs}\label{APP:00A}
\subsection*{Proof of Theorem~\ref{THM:NPCov-from-path-weight}}
If we set $P=\{x, y\}$ so that $|P|=2$ and $\bar{P}=V\backslash \{x,y\}$ then it
follows from equation (\ref{EQN:path.weight1}) that
\begin{eqnarray}
  \nonumber
  \w(\lp x, y\rp{}, \Sigma)
  &=& (-1)\; |\Sigma_{PP}|\; \kappa_{xy}\\
  \nonumber
  &=& (-1)\; |\Sigma_{PP\given\bar{P}}|\; \kappa_{xy}\; \frac{|\Sigma_{PP}|}{|\Sigma_{PP\given\bar{P}}|}\\
  \label{EQN:thm.two.comp.01}
  &=& \frac{-\kappa_{xy}}{|\Kita_{PP}|}\; \frac{|\Sigma_{PP}|}{|\Sigma_{PP\given\bar{P}}|}
\end{eqnarray}
where in equation (\ref{EQN:thm.two.comp.01}) we have used the fact that
$\Kita_{PP}=\Sigma_{PP\given \bar{P}}^{-1}$. We note that in equation (\ref{EQN:thm.two.comp.01})
we have
\begin{eqnarray*}
  \frac{-\kappa_{xy}}{|\Kita_{PP}|}=\{\Sigma_{PP\given\bar{P}}\}_{xy}=\sigma_{xy\given V\backslash\{x,y\}},
\end{eqnarray*}
and, furthermore, it follows from the definition of the vector alienation coefficient and
equation (\ref{EQN:vector.correlation}) that
\begin{eqnarray*}
\frac{|\Sigma_{PP}|}{|\Sigma_{PP\given\bar{P}}|}
=\left(\frac{|\Sigma_{P\cup\bar{P} P\cup\bar{P}}|}{|\Sigma_{PP}||\Sigma_{\bar{P}\bar{P}}|}\right)^{-1}
=\frac{1}{\lambda_{\vc{P}{\bar{P}}}}
=\frac{1}{1-\rho^{2}_{\vc{P}{\bar{P}}}}.
\end{eqnarray*}
Hence, (\ref{EQN:thm.two.comp.01}) can be written in the form
\begin{eqnarray*}
 \w(\lp x, y\rp{}, \Sigma) = \frac{\sigma_{xy.\bar{P}}}{1-\rho^{2}_{\vc{P}{\bar{P}}}},
\end{eqnarray*}
as required.

\subsection*{Proof of Theorem~\ref{THM:path.update}}
Let $A=V\backslash Q$. If $\Kita$ implies the graph $\G=(V, \E)$, then $\Sigma_{AA\given Q}^{-1}=\Kita_{AA}$
implies the subgraph $\G_{A}=(A, \E_{A})$. Hence, if $\Pt\in\Ptxy_{xy}$ is a path between $x$ and $y$ in
$\G$ such that $V(\Pt)\subseteq A$ then $\Pt$ is also a path between $x$ and $y$ in $\G_{A}$ and it makes
sense to compute the weight of $\Pt$ with respect to the distribution of $X_{A}|X_{Q}$, that is
$\w(\Pt, \Sigma_{AA\given Q})$. More specifically, it follows from equation (\ref{EQN:path.weight1}) and
(\ref{EQN:path.weight2}) that
\begin{eqnarray*}
  \w(\Pt, \Sigma_{V\backslash QV\backslash Q\given Q})
  &=& (-1)^{|P|+1}\;  |\Sigma_{PP\given Q}|\; \prod_{\{u,v\}\in \E(\Pt)} \kappa_{uv}.
\end{eqnarray*}
and an immediate consequence of Theorem~\ref{THM:jones.west} is that
\begin{eqnarray}\label{EQN:sigmaxy.c1.new}
\sigma_{xy\given Q}
    &=&\sum_{\Pt\in \Ptxy_{xy}; V(\Pt)\subseteq A} \w(\Pt, \Sigma_{AA\given Q})
\end{eqnarray}
where, for $\Pt\in \Ptxy_{xy}$ with $V(\Pt)\subseteq A$,
\begin{eqnarray*}
    \w(\Pt, \Sigma_{AA\given Q})&=& (-1)^{|P|+1}\;  |\Sigma_{PP\given Q}|\; \prod_{\{i,j\}\in \E(\Pt)}\kappa_{ij}.
\end{eqnarray*}

If we divide both sides of equation (\ref{EQN:sigmaxy.c1.new}) by
\begin{eqnarray*}
  1-\rho^{2}_{\vc{xy}{Q}}
  =\frac{|\Sigma_{QQ\given \{x, y\}}|}{|\Sigma_{QQ}|},
\end{eqnarray*}
then we obtain
\begin{eqnarray*}
\ew_{\ewsub{xy}{Q}}
    &=&\sum_{\Pt\in \Ptxy_{xy}; V(\Pt)\subseteq A} \w(\Pt, \Sigma_{AA\given Q})\frac{|\Sigma_{QQ}|}{|\Sigma_{QQ\given \{x, y\}}|},
\end{eqnarray*}
where
\begin{eqnarray*}
\w(\Pt, \Sigma_{AA\given Q})\frac{|\Sigma_{QQ}|}{|\Sigma_{QQ\given \{x, y\}}|}
&=&
(-1)^{|P|+1}\;  \frac{|\Sigma_{PP\given Q}||\Sigma_{QQ}|}{|\Sigma_{QQ\given \{x, y\}}|} \; \prod_{\{i,j\}\in \E(\Pt)}\kappa_{ij}\\
&=&
(-1)^{|P|+1}\;  \frac{|\Sigma_{QQ\given P}||\Sigma_{PP}|}{|\Sigma_{QQ\given \{x, y\}}|} \; \prod_{\{i,j\}\in \E(\Pt)}\kappa_{ij}\\
&=& \w(\Pt, \Sigma)\times\frac{|\Sigma_{QQ\given P}|}{|\Sigma_{QQ\given \{x,y\}}|}\\
&=&  \w(\Pt, \Sigma)\times(1-\rho^{2}_{\cvc{P\backslash\{x,y\}}{Q}{\{x,y\}}}),
\end{eqnarray*}
as required.
\section{Limited-order networked partial covariance decomposition on four vertices}\label{APP:00C}
\begin{figure}
\begin{center}
\begin{figurepdf}
\includegraphics[scale=.8]{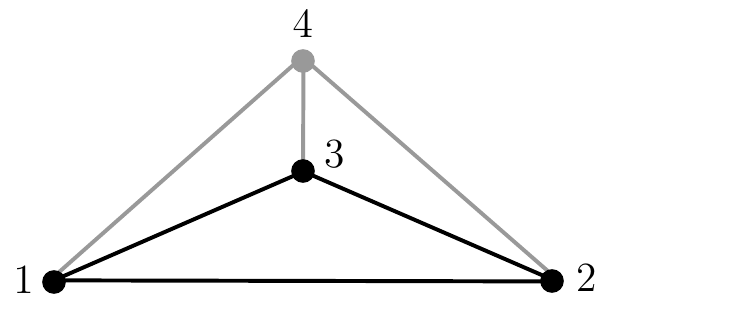}
\end{figurepdf}
\caption{Complete graph on four vertices: the grey part highlights the connection of vertex 4 with the rest of the graph.}\label{FIG:complete.4}
\end{center}
\end{figure}

For the graph in Fig.~\ref{FIG:complete.4} we focus on the decomposition of the
covariance $\sigma_{12}$, i.e. on all the paths between vertices $1$ and $2$.
It follows from Theorem~\ref{THM:jones.west} that $\sigma_{12}$ can be computed
as the sum of the five path weights that are given in Table~\ref{TAB:table-weights1},
where we use the suppressed notation $\Sigma_{12}$ to denote $\Sigma_{\{1,2\}\{1,2\}}$.
We remark that this example also covers the case where the graph is not complete
because it is sufficient to recall that the corresponding entry of the concentration
matrix is equal to 0 and, consequently, the same is true for every path involving
such edges.
\begin{table}\caption{Weights of the paths between vertices $1$ and $2$ in the graph of Fig.~\ref{FIG:complete.4}}\label{TAB:table-weights1}
\begin{center}
  \begin{tabular}{|l|lll|}
    \hline
    \multicolumn{1}{|c|}{Path}  & \multicolumn{3}{c|}{Path weight}\\
    \hline
    $1\edge\ 2$ & $\ew_{\ewsub{12}{34}}$ &$=$ & $-\kappa_{12}$  $|\Sigma_{12}|$\\
    $1\edge\ 3\edge\ 2$ & $\w(\lp 1,3,2 \rp, \Sigma)$ &$=$ & $+\kappa_{13}\;\kappa_{32}$ $|\Sigma_{123}|$\\
    $1\edge\ 4\edge\ 3\edge\ 2$ & $\w(\lp 1,4,3,2 \rp, \Sigma)$ &$=$ &$-\kappa_{14}\;\kappa_{43}\;\kappa_{32}$ $|\Sigma|$\\
    $1\edge\ 3\edge\ 4\edge\ 2$  & $\w(\lp 1,3,4,2\rp, \Sigma)$ & $=$ &$-\kappa_{13}\;\kappa_{34}\;\kappa_{42}$ $|\Sigma|$\\
    $1\edge\ 4\edge\ 2$   & $\w(\lp 1,4,2\rp, \Sigma)$ & $=$ &$+\kappa_{14}\;\kappa_{42}$ $|\Sigma_{124}|$\\
    \hline
  \end{tabular}
\end{center}
\end{table}

Consider now the case where we marginalize over $X_{4}$ so that $Q=\{3\}$. If we
write $\Sigma^{-1}_{Q\cup\{x,y\}Q\cup\{x,y\}}=\Sigma_{123}^{-1}=\{\kappa^{*}_{ij}\}_{i,j\in \{1,2,3\}}$,
then the weights associated with the paths between $1$ and $2$ in the subgraph of the graph
in Fig.~\ref{FIG:complete.4} induced by $\{1,2,3\}$ are given in Table~\ref{TAB:table-weights2}.
\begin{table}
\caption{Weights of the paths between vertices $1$ and $2$ in the subgraph of
         the graph in Fig.~\ref{FIG:complete.4} induced by $\{1,2,3\}$, after
         marginalization over variable $X_{4}$.}\label{TAB:table-weights2}
\begin{center}
  \begin{tabular}{|l|lll|}
    \hline
    \multicolumn{1}{|c|}{Path}  & \multicolumn{3}{c|}{Path weight}\\
    \hline
    $1\edge\ 2$ & $ \ew_{\ewsub{12}{3}}$ & $=$ &$-\kappa^{*}_{12}$  $|\Sigma_{12}|$\\
    $1\edge\ 3\edge\ 2$ & $\w(\lp 1,3,2 \rp, \Sigma_{123})$ &$=$ & $+\kappa^{*}_{13}\;\kappa^{*}_{32}$ $|\Sigma_{123}|$\\
    \hline
  \end{tabular}
\end{center}
\end{table}
It follows from Theorem~\ref{THM:path.update} that
\begin{eqnarray*}
  \ew_{\ewsub{12}{3}}
  &=& \ew_{\ewsub{12}{3}}\\
  &+& \w(\lp 1,4,2\rp, \Sigma)\times (1-\rho^{2}_{34\given 12})
\end{eqnarray*}

If we exploit the fact that the sum of the five path weights in Table~\ref{TAB:table-weights1}
is equal to the sum of the two path weights in Table~\ref{TAB:table-weights2}, i.e. equal to
$\sigma_{12}$, then we can decompose the weight of the path $\lp 1, 3,  2\rp$, relative to
$\Xbf_{\{1,2,3\}}$, as follows
\begin{eqnarray*}
  \w(\lp 1,3,2 \rp, \Sigma_{123})
  &=& \w(\lp 1,3,2 \rp, \Sigma)   \\
  &+& \w(\lp 1,4,3,2 \rp, \Sigma)\\
  &+& \w(\lp 1,3,4,2 \rp, \Sigma)\\
  &+&\w(\lp 1,4,2 \rp, \Sigma)\times \rho^{2}_{34\given 12}.
\end{eqnarray*}

We can conclude that the weight, relative to $\Xbf_{\{1,2,3\}}$, of the path
$\lp 1, 2\rp$ can be obtained by adding to the weight, relative to
$\Xbf_{\{1,2,3,4\}}$, of the path $\lp 1, 2\rp$ the proportion
$(1-\rho^{2}_{34\given 12})$ of the weight of path $\lp 1, 4, 2\rp$; note that
$(1-\rho^{2}_{\vc{34}{12}})=1$ if the edge $\{3, 4\}$ does not belong to the
graph. In addition, the weight, relative to $\Xbf_{\{1,2,3\}}$, of the path
$\lp 1, 3, 2\rp$ can be be obtained by adding to the weight, relative to
$\Xbf_{\{1,2,3,4\}}$, of the path $\lp 1,3,2\rp$ the proportion
$\rho^{2}_{34\given 12}$ of the weight of path
$\lp 1, 4, 2\rp$, with $\rho^{2}_{\vc{34}{12}}=0$ if $\{3, 4\}$ does not belong
to the graph, and, furthermore, the weights of all the remaining paths between
$1$ and $2$ in the graph which involve the vertex 3.
%

%
\bibliographystyle{chicago}
\bibliography{pw-ref}
\appendix
\end{document}